\documentclass[conference]{IEEEtran}
\usepackage{float}
\usepackage{mdwmath}
\usepackage{mdwtab}
\usepackage{amsfonts}
\usepackage{algorithm}
\usepackage{algpseudocode}
\usepackage{lipsum}
\usepackage{cite}
\usepackage{algorithmicx}
\usepackage{algpseudocode}
\ifCLASSINFOpdf
  \usepackage[pdftex]{graphicx}
\else
\fi
\usepackage[cmex10]{amsmath}

\usepackage{mdwmath}
\usepackage{mdwtab}

\usepackage{graphicx}
\usepackage{algorithm}
\usepackage{algpseudocode}
\usepackage{lipsum}
\usepackage{amsmath}
\usepackage{amsfonts}
\usepackage{verbatim}
\usepackage{verbatim}
\usepackage{algpseudocode}
\usepackage{multicol}
\usepackage{filecontents}

\usepackage{esvect}
\usepackage{relsize}
\usepackage{booktabs}
\usepackage{multirow}
\usepackage{siunitx}

\newcommand{\qed}{\nobreak \ifvmode \relax \else
      \ifdim\lastskip<1.5em \hskip-\lastskip
      \hskip1.5em plus0em minus0.5em \fi \nobreak
      \vrule height0.75em width0.5em depth0.25em\fi}

 \usepackage{tikz}
        \usepackage{tikz-qtree}
        \usetikzlibrary{matrix}

\usepackage{subcaption}
\usepackage{kantlipsum} 
\usepackage{mwe} 
\usepackage{tablefootnote}

\DeclareCaptionLabelSeparator{periodspace}{.\quad}
\usepackage{caption}
\begin{document}
%
\title{Anomaly Detection for Industrial Control Networks using Machine Learning with the help from the Inter-Arrival Curves}
%
%
%

\author{\IEEEauthorblockN{
Basem~AL-Madani,
Ahmad~Shawahna, and
Mohammad~Qureshi
}
\IEEEauthorblockA{Department of Computer Engineering\\
King Fahd University of Petroleum and Minerals, Dhahran-31261, KSA\\ 
\{mbasem, g201206920, qaziumer\}@kfupm.edu.sa\\
}
}


\maketitle


\begin{abstract}

Industrial Control Networks (ICN) such as Supervisory Control and Data Acquisition (SCADA) systems are widely used in industries for monitoring and controlling physical processes. These industries include power generation and supply, gas and oil production and delivery, water and waste management, telecommunication and transport facilities. The integration of internet exposes these systems to cyber threats. The consequences of compromised ICN are determine for a country economic and functional sustainability. Therefore, enforcing security and ensuring correctness operation became one of the biggest concerns for Industrial Control Systems (ICS), and need to be addressed. In this paper, we propose an anomaly detection approach for ICN using the physical properties of the system. We have developed operational baseline of electricity generation process and reduced the feature set using greedy and genetic feature selection algorithms. The classification is done based on Support Vector Machine (SVM), k-Nearest Neighbor (k-NN), and C4.5 decision tree with the help from the inter-arrival curves. The results show that the proposed approach successfully detects anomalies with a high degree of accuracy. In addition, they proved that SVM and C4.5 produces accurate results even for high sensitivity attacks when they used with the inter-arrival curves. As compared to this, k-NN is unable to produce good results for low and medium sensitivity attacks test cases.

\end{abstract}


\IEEEpeerreviewmaketitle

\section{Introduction}

Supervisory Control and Data Acquisition (SCADA) is a type of Industrial Control Systems (ICS). SCADA is used to monitor, supervise, and control the industrial processes. It is in use for several decades and plays an important role in operation of many industries. SCADA is responsible for monitoring and controlling the industrial process to ensure a continuous operation, and the safety of the plant and human beings. SCADA system involves the data transfer between a central SCADA control center, Remote Terminal Units (RTU), Programmable Logic Control (PLC), and operator terminals. Note that tracing the transferred data helps in providing information about the status of the monitored system.

The evolution of SCADA system proceeded in three generation. In the first generation, SCADA systems were isolated network and have neither connection with computer networks nor other standard systems. They used proprietary protocols for communication with a very limited connectivity. The communication capabilities were enhanced in the second generation through the integration with local area network. The current SCADA system, also known as Internet-based SCADA, uses standard networking protocols commercial products. Internet and Enterprise Resource Planning (ERP) systems are fully integrated with the current generation SCADA systems.

The security of the industrial control systems is critical for plant operations and safety. Any disruption in the services monitored and supervised through ICS has adverse consequences. Typical services include power generation and distribution, water treatment plant, and nuclear facilities. Traditionally, SCADA systems are believed to be secure from external threats as they are not linked to external networks and used proprietary protocols. However, the introduction of Internet and commercial applications have exposed SCADA systems to security threats although that they provided numerous benefits to SCADA systems \cite{teixeira2011cyber}. Furthermore, the most prevalent SCADA communication protocols have inherent security weaknesses due to the lack of authentication, integrity check, and encryption \cite{sommestad2010scada}. These weaknesses can be exploited to gain unauthorized access, overwrite running configuration, and execute Denial of Service (DoS) and Man-In-The-Middle (MITM) attacks. The discovery of the Stuxnet malware, which was used to attack the nuclear plant in Iran, also exposes these security weaknesses. In this incident, hardcoded authentication credentials were used to compromise connection between operator’s station, PLC, and monitored equipments.

The security measures of data network are usually applied to ICNs which includes firewalls, Intrusion Detection System (IDS), and Anomaly Detection System (ADS). However, the data network and ICN are very different from each other. Thus, there is a need for the development of special security product for industrial control networks. Generally, industrial control networks have strict timing requirements and regular sequence of operations \cite{galloway2013introduction}. This makes ICNs more predictable which can be exploited in the development of ADS. It has also been believed by many researchers that ADS is more favorable than rule-based detection methods \cite{bhuyan2014network}. However, the problem with ADS is the high number of false positives. In addition, as anomaly detection is done by developing normal profile of operations, a patient attacker can alter the normal profile by long undetected attacks. These modifications make the attacks acceptable to ADS.

The proposed ADS tackles both of the previously mentioned challenges using sliding window trim mean and developed equation for determining the thresholds. We have considered attacks that can corrupt the normal command and response messages between the controlling station and the controlled process. In other words, we have assumed that an attacker can compromise command and response messages. This assumption is valid due to the lack of authentication, integrity checking, and replay attack detection in prevalent SCADA communication protocols. The compromising of these messages includes changing the values of the parameters in response or command. This corruption can disturb the operation of the process and force an equipment to operate in a danger zone. Preventing the system from operate in a danger zone is one of proposed ADS goals.

The remainder of this paper is organized as follows. Section II discusses the vulnerabilities in industrial control systems. The related work and introduction to power generation process are provided in section III. The implementation details of this work is discussed in section IV. The experimental work is deliberated in section V. The accuracy, false positive, and attack detection rate results and their analysis are presented in section VI. Section VII concludes the paper.

\section{Vulnerabilities In Industrial Control Systems}

The integration of the latest information and communication technologies into ICS enhances its functionalities and connectivity. Nowadays, industrial control systems are tightly integrated with Internet and use the same network used by other TCP/IP devices. This integration exposes these systems to all vulnerabilities of the information and communication technologies \cite{cagalaban2010improving}. In addition, ICS used to have proprietary software but current ICS uses commercial software with known vulnerabilities such as Microsoft Windows and SQL. Beside these vulnerabilities, communication protocols and infrastructure of ICS lack basic security mechanism such as authentication and integrity checks. These vulnerabilities expose ICS to multitude of threats which can result in catastrophic events. In this section, we discuss security vulnerabilities in ICS infrastructure and communication protocols and their impacts when exploited.

\subsection{ICS Communication Protocol Vulnerabilities}

The communication protocols of ICS are designed to ensure availability, reliability, and timely execution of tasks. Most industrial control communication protocols are designed and developed when the security was not the prime concern. This results in the inherent security vulnerabilities in these protocols. Basic security features such as authentication, integrity check, and time checks are missing in ICS communication protocols. Exploiting these vulnerabilities results in compromising the system and providing complete control to the attacker. The authors in \cite{fovino2010taxonomy} presented the vulnerabilities of two major communication protocols for industrial control system; MODBUS and DNP 3.0. Although the vulnerabilities are specific to these protocols, they are common among other communication protocols as well. These vulnerabilities can be exploited to compromise integrity and safety as well as launch DoS and man in the middle attacks. Several attacks on industrial control systems have been reported in \cite{nicholson2012scada}.

\subsection{Supervisory Layer Vulnerabilities}

An operator system is used to monitor and supervise industrial process. A compromised operator station has proven to be lethal. This is one of entry points for an attacker to launch an attack. The followings are possible threats and vulnerabilities in an operator system: i) The operator workstation has evolved into PCs and based on Windows Platform. It is known that Windows platform has high number of known vulnerabilities and unpatched PC can be easily exploited \cite{zhu2011taxonomy}.  ii) The system can also be compromised due to weak passwords, installation of unauthorized software, connection to internet and absence or obsolete Anti-Virus. iii) The use of commodity hardware and software solution such as Microsoft Windows, TCP/IP networking, and SQL database exposes industrial control systems to same vulnerabilities of these products \cite{rautmare2011scada}. This adoption is mainly due to their low cost, high availability, and high connectivity requirements.

\subsection{Field Layer Vulnerabilities}

Remote terminal units are responsible for managing field devices. On the other hand, programmable logic control is commonly used to control filed devices and gather telemetric data. PLCs are programmed by using software such as STEP 7 by Siemens. The potential for reprogramming an RTU or PLC by accessing the polling/communication circuit exists. The attacker can replace the valid configuration file by malicious file. This threat could be exploited in case of PLC without providing the means for authentication or having hardcoded username and password in the firmware.


\section{Related Work}

The security of SCADA system can be achieved by two approaches. The first approach considers securing the perimeter of SCADA network by firewalls, IDS and host based IDS, or anti-virus. However, the previously mentioned technologies are insufficient in securing SCADA systems due to the unique nature of SCADA systems gaps \cite{johnson2010survey}. On the other hand, the second approach is to develop profiles of normal operations and detect intrusions using these profiles \cite{barbosa2010intrusion}. The second approach can be further divided on the basis of feature sets used in developing these normal operations' profiles. The feature sets can be related to protocol parameters, network traffic patterns, or measurements from physical processes such as pressure, speed, and power.

An anomaly based IDS to detect intrusion by establishing network traffic flow model and relationship between them is proposed in \cite{barbosa2010intrusion}. The proposed IDS is intended to automatically generates flow models and detect anomalies in network traffic that violate the established traffic flow model. It is potential to generate all possible traffic flow in SCADA network because of the limited number of devices, protocols and regular communication patterns. The implementation and results of proposed IDS is presented in \cite{barbosa2012difficulties} along with the challenges in modeling SCADA traffic. The authors used traditional five tuple protocol number, source and destination IP addresses, and port numbers to generate network traffic flow information. They used invariants for analyzing the SCADA network traffic. These invariants include diurnal pattern of activities which describe the network traffic based on time and weekdays. In addition, they include lognormal connection sizes and heavy-tail distributions which describe connection size distribution and self-similarity. The results shown that SCADA network have regular and periodic patterns in nature, and both lognormal and heavy-tail do not provide good fit for the SCADA connection size. The authors mentioned that existing traffic models cannot be easily applied to SCADA traffic.

A neural network based intrusion detection system is proposed in \cite{gao2010scada} for SCADA lab system for water tank storage. The proposed IDS uses four features as an input parameters to represent the physical state of the system; water level, command response frequency, mode of operation, and water tank pump state. The proposed attack model consists of command and response injection and DoS attacks which affect the integrity and the availability aspect of the SCADA network, respectively. The results shown that the proposed IDS is able to achieve high accuracy for detecting DoS attacks and command and response injection attacks but with high percentage of false positive in detecting replay attacks. The proposed IDS approach is designed for water storage tank with a simple process of monitoring. In such away, the IDS accuracy may degraded if it is applied to more complex system.

Alavaro et al. \cite{cardenas2011attacks} presented that incorporating physical system knowledge enables the identification of critical sensors and the attacks on them. They modeled Tennessee-Eastman Process Control System (TE-PCS) which is responsible for controlling a chemical reaction with primary objective of maintaining pressure around 3000 kPa. The proposed detection method is based on change detection with the use of cumulative sum which detects the change between two hypotheses in minimum possible time. Three types of attacks, namely surge, bias and geometric, are modeled and launched on different sensors in the network. They found that denial of service attacks do not force the control system to operate in unsafe condition. However, the attacks on integrity of sensor force the change in pressure beyond safety limits. An automatic detection module is also proposed which replaces the sensor measurements by estimated measurements using linear model of system when an intrusion is detected.

In \cite{garitano2011review}, two approaches to enhance the accuracy of anomaly detection in SCADA systems have been discussed. The first approach is based on n-gram technique which records the normal operation of the SCADA network. The second approach is based on the invariant induction which establishes mathematical relationship between different data readings and uses these relationships to model the normal operation of the electric network. It detects the corrupted measurement due to fault or attack by evaluating it against mathematical model. The authors used load flow program to record real and reactive power flow measurements of six buses electric network of varying load for one year period. They introduced 1 to 44 random errors in the calculated data files. They concluded that invariant induction has better overall performance, while n-gram is better in detecting corrupt files. It is also suggested to use correlation of two or more anomaly detection techniques for further enhanced accuracy.

A Bayesian based correlation technique is used to extended the work discussed in \cite{garitano2011review} by correlate the output of two IDS \cite{sridhar2012cyber}. The authors used invariant induction and artificial ant approach based IDS to reduce false positives. They defined three invariant checkers for profiling and anomaly detection, a linear invariant checker, range checker, and a bus-zero-sum checker. The artificial ant approach is adopted by clustering real and reactive powers to define normal operation of electric network by using Bayesian based correlation.

An anomaly detection system based on data rough set theory \cite{ chikalov2013rough} is mentioned in \cite{ pisicua2010rough} for securing the electric power system. The proposed approach profiles the normal operation by extracting rules from the normal operation and compare incoming measurements from remote terminal units with the normal profile. The authors have reduced the number of rules for anomaly detection which makes the rule dynamic and less resources intensive. The experimentation is done on the six bus power system, the data set consists of reading 45 tests with 57 measurement values. The errors are introduced on bus 4 and bus 6; the authors only introduce switch sign error in the dataset. The rules for detection anomalies are based on the power flow between the buses on the electrical network.

Similar to data networks, the normal profiling can also be done by using features from network protocol, traffic patterns, and client server response. There is considerable amount of research done for ICN security based on these concepts. In \cite{zhu2010scada}, intrusion detection systems that modeled the specification of the MODBUS/TCP are discussed. The authors in \cite{goldenberg2013accurate} exploited the static nature of control system network topology and network traffic pattern. They proposed three techniques to detect intrusions on the control network. The first approach based on the protocol specification of the fields in the request and reply messages of MODBUS/TCP. They have developed specification for function codes, exception codes, protocol identifiers and cross field relationship specification such as the relationship between the length and the function code. The second approach exploits the predictable communication pattern in control systems, snort \cite{chakrabarti2010study} is used to detect suspicious communication patter. The snort rules are developed based on the modeled communication pattern in MODBUS/TCP based network. In the third approach, they used heuristics to learn the availability of the server and clients on the system which helps in identifying rogue devices and changing network services.

Hui Lin et al. \cite{lin2013adapting} proposed an intrusion detection system for DNP 3 protocol which uses Bro IDS \cite{mehra2012brief}. Bro IDS is a network analysis framework. The proposed system modified Bro and consists of three main components; a network parser, an event handler, and a policy script interpreter which has Protocol Validation Policy (PVP). The network packet parser is responsible for decoding incoming packet. The event handler acts as an interface between DNP3 parser and the policy script interpreter. Event handlers are specified for each type of data field of DNP3 protocol. PVP is responsible to look for protocol specification violations, it performs two types of validation; inter-packet validation, and intra-packet validation. In inter-packet validation, it looks for anomalous communication patterns such as unmatched request/reply. The system can keep the history of states from parsed network packets. This type of validation helps in detecting DoS and replay attacks. The intra-packet validation validates the dependencies between different data fields to detect malformed packets which causes DoS, an anomalous packet has mismatch of length field and actual length of the real pay load. The authors have conducted robustness and throughput testing in the experiments but do not test the proposed system for malicious activity. 

The authors in \cite{salem2016anomaly} proposed a sliding window based anomaly detection framework. The proposed framework uses a specific form of arrival curves, denoted as inter-arrival curves, for analyzing discrete event traces, which helps in performing a better detection of abnormal behavior of the real-time embedded systems. The proposed framework uses the input trace to calculate two special types of inter-arrival curves; maximum inter-arrival curve ($C_{max}$), and minimum inter-arrival curve ($C_{min}$). During classification phase, the classifier used the obtained features $C_{max}$ and $C_{min}$ from the tested trace to evaluate their conformance to the corresponding features obtained from the set of known traces in the training model. The classification framework considers the behavior of an event to be anomalous when its $C_{max}$, $C_{min}$, or both are anomalous. The proposed approach only considers events occurring within a defined time step, which means that their exact occurrence time are neglected. This implies that anomalies affecting events timing are not detected by inter-arrival curves based framework. In addition, min and max inter-arrival curves are unable to detect anomalies which have a slight perturbation to the event trace.

\section{Proposed Method}

This section discusses the developed anomaly detection technique for industrial control system in power generation sector. In this area, most of the work has been done in security of power distribution networks only. The proposed anomaly detection system uses artificial intelligence techniques such as support vector machine, k-nearest neighbor, and C4.5 decision tree for power generation process. In addition, it uses the inter-arrival curves for better detection of abnormal behavior of industrial control systems. 

Industrial control systems are often implemented to perform a recurrent behavior. Thus, the generated event traces by these systems are expected to reflect such a recurrence. This allows for analyzing the system behavior with respect to an expected behavior. Moreover, ICS equipments operate in predefined ranges, such kind of operations are easily to profile. The proposed anomaly detection system has a dual detection frameworks that detects abnormal behavior by analyzing both the equipments readings with their thresholds as well as the discrete event traces generated by the industrial control system.

\subsection{Threshold-Based Anomaly Detection}

In power generation process, the equipments operate in predefined ranges, such kind of operations are easily to profile. An anomaly is defined as any reading which violates these normal profiles. The following subsections illustrate the classification techniques considered in the proposed anomaly detection system for ICS.

\subsubsection{Support Vector Machine (SVM)}
\hfill \par
A support vector machine is a supervised machine learning system that has been widely employed for intrusion detection \cite{gryllias2012support}. It classifies the input vectors into two-classes or multi-classes. In this research, we are considering two classes; normal class and anomalous class. The normal class is represented by +1 while the anomalous class is represented by -1. The basic concept of SVM is to find a margin between the two classes in a hyperplane. In addition, its main objective is to maximize the margin which provides better generalization ability \cite{durgesh2010data}. The generalization ability is defined as classifier accuracy for current and future data. The set of training vectors that belongs to the considered two separate classes, normal and attack classes, can be represented using Equation~\ref{setTrainingVector}.

\begin{equation}
  S = \bigg\{\big(x^1,~y^1\big),~\big(x^2,~y^2\big),~....~,~\big(x^l,~y^l\big)\bigg\}
  \label{setTrainingVector}
\end{equation}

where, $x^i \in \mathbb{R}^n$ is a vector in the two classes, $i = 1, 2, ..., l$, and $y \in \mathbb{R}^l$ is an indicator vector such that $y^i \in \big[+1, ~-1\big]$

The optimal separation hyperplane for the data points is defined using Equation~\ref{optimalHyperplane}. The closest data points to the margin are called support vectors. Note that if the training data can be linearly separated, then there exists a pair $(\vv{w},~b)$ for the data points $x^i$ as shown in Equation~\ref{normalequation} and Equation~\ref{anomulousequation}.

\begin{equation}
  \vv{w}^T x + b = 0
  \label{optimalHyperplane}
\end{equation}

\begin{equation}
  \vv{w}^Tx^i + b \geq +1,~~~x^i \in Normal
  \label{normalequation}
\end{equation}

\begin{equation}
  \vv{w}^Tx^i + b \leq -1,~~~x^i \in Anomalous
  \label{anomulousequation}
\end{equation}

where, $\vv{w}$ is the weight vector, and $b$ is the scalar (bias).

Therefore, the classification problem using the training set can be estimated using function $f: \mathbb{R}^n \mapsto \{\pm1\}$ as represented in Equation~\ref{functionSeperation}.

\begin{equation}
  f(x) = \begin{cases} 
      +1 &,~\vv{w}^Tx + b \geq +1 \\
      -1 &,~\vv{w}^Tx + b \leq -1
   \end{cases}
  \label{functionSeperation}
\end{equation}

The optimal separating hyperplane that provide the largest margin between two classes can be found by minimizing squared norm of separating hyperplane \cite{chang2011libsvm}. Hence, the hyperplane that optimally separates the data is the one that minimizes $\Phi$ which can be calculated using Equation~\ref{phiequation}.

\begin{equation}
 \Phi = \frac{{~\|\vv{w}\|}^2}{2}
  \label{phiequation}
\end{equation}

If the data points are not linearly separable, the margin may be negative. In this case, SVM employ the concept of soft margin which relax the requirement of Equation~\ref{normalequation} and Equation~\ref{anomulousequation}. The second approach is to use the kernel which linearize the data points to make them separable.

We utilize SVM in this research due to its good generalization ability of the learning model. This implies that good accuracy can be achieved even with relatively smaller training datasets with SVM. Another advantage of SVM is its ability to handle a large number of features. In addition, SVM also ensures high accuracy for classification of future data from the same allotment to which the training data belongs \cite{durgesh2010data}. Moreover, SVMs are free from the problem of over-fitting. Experimental results obtained confirm high accuracy of attack detection and insignificant false positive rates for the proposed approach.

\subsubsection{K-Nearest Neighbor (k-NN)}
\hfill \par
The k-nearest neighbor algorithm is a well-known machine learning methodology characterized by a good performance and a short training period \cite{suguna2010improved}. K-NN is considered as a type of the non-parametric classifiers and special case of the instance-based classifiers \cite{olvera2010review}. The working principle of the k-NN algorithm based on the assumption that instances of the same class generally lie in close to each other. The label of the unknown instance can be determined by examining the class of the closest neighbors. K-NN classifier is trained with the labelled training objects. Then, the classification of the test object is done by computing the approximate distance or the similarity between the test object and the training objects. The objective of the distance metric is to minimize the distance between two points of same class, while maximize the distance between different classes. The distance determines the nearest neighbor to the test object. Subsequently, the class of the k-nearest neighbor is assigned to the test object. The most common distance calculation methods are Euclidean and Manhattan \cite{sinwar2014study} as shown in Equation~\ref{Euclidean} and Equation~\ref{Manhattan}, respectively.

\begin{equation}
 d(x, y) = \sqrt{\sum_{i=0}^{n} {(x_i - y_i)}^2}
  \label{Euclidean}
\end{equation}

\begin{equation}
 d(x, y) = \sum_{i=0}^{n} |x_i - y_i|
  \label{Manhattan}
\end{equation}

where, $x$ is the vector to be classified, $y$ is the trained vector, and $n$ is the dimensional space.

Furthermore, other distance calculation methods could be used for specific problems. Euclidean distance method is used in the proposed ADS. To enhance the accuracy of the k-NN, weighting scheme is used to influence the distance measurement and voting of each instance. In the k-NN, the value of $k$ is an important parameter since it determines the number of neighbors to be included in the nearest neighbor list. Smaller value of $k$ makes the scheme sensitive to noise whereas larger value will include objects from other classes. In other words, the accurate choice of $k$ enhances the performance of the algorithm. The optimal value of the $k$ can be obtained by cross-validation technique \cite{hardle2012smoothing}. In this research, the value of $k$ is set to be equal to 1 according to the results of the cross-validation which is also appropriate for small datasets \cite{dasgupta2014optimal}.

The main advantage of k-NN methods is their simplicity and lack of parametric assumptions. These methods perform surprisingly well, especially when each class is characterized by multiple combinations of predictor values as in the case of the dataset used in this research work. For instance, an anomalous behavior can be characterized by one, three or five parameters in the used dataset.

\subsubsection{C4.5 Decision Tree}
\hfill \par
C4.5 \cite{chauhan2013implementation} is one the most popular tree classifiers. C4.5 algorithm forms a decision tree to describe the classifier and uses the decision tree to produce the rule set. Due to the distributed nature of information about a single class in a tree, it may be difficult to interpret complex decision trees. As such, C4.5 uses groups of if-then rules for each class to classify a case \cite{farid2010attacks}. The first rule that matches the conditions outlined by a case classifies that case. In case no matching rule occurs, C4.5 assigns a default class to the case.

The initial, unpruned, decision tree forms the C4.5 rule sets. Each path from the root of the tree to a leaf becomes a prototype rule whose conditions are the outcomes along the path and whose class is the label of the leaf. To simplify the rule, the consequence of discarding each condition is studied. Dropping a condition may result in rise the instances in the instance set of cases covered by the rule, and the number of errors of cases that do not belong to the class nominated by the rule, and may lower the pessimistic error rate. The optimal pessimistic error rate is found by dropping conditions using a hill-climbing algorithm \cite{borghoff2010hill}. To finalize the process, C4.5 picks a subset of simplified rules for each class and sorts them to minimize the error on the training cases. In addition, it chooses a default class. As a result, the obtained rule set usually has far fewer rules than the number of leaves on the unpruned decision tree.

\subsection{Event-Based Anomaly Detection}

The proposed anomaly detection framework uses a specific form of arrival curves \cite{moy2010arrival}, denoted as inter-arrival curves \cite{salem2016anomaly}, for analyzing discrete event traces, which helps in performing a better detection of abnormal behavior of the industrial control systems. The proposed framework uses the input event trace to calculate two special types of inter-arrival curves; maximum inter-arrival curve ($C_{max}$) and minimum inter-arrival curve ($C_{min}$). $C_{max}$ provides the maximum counts for occurrences of event $e$ within the sliding windows. However, $C_{min}$ represents the corresponding minimum counts of event $e$ within the sliding window. Note that the sliding window always start with the considered event $e$ to optimize the computation of inter-arrival curves, and the length of the sliding windows varies between one and the maximum window size threshold ($W_\delta$). $C_{min}$ and $C_{max}$ can be calculated using Equation~\ref{cmin} and Equation~\ref{cmax}, respectively. The minimum and the maximum inter-arrival curves can be affected by abnormal behavior. Therefore, the proposed classification framework uses these curves to detect and quantify such anomalies.

\begin{equation}
  C_{min}(T)  = \operatornamewithlimits{argmin}\limits_{w = 1, 2, ..., W_\delta}\operatorname{C}(T, w)
  \label{cmin}
\end{equation}

\begin{equation}
  C_{max}(T)  = \operatornamewithlimits{argmax}\limits_{w = 1, 2, ..., W_\delta}\operatorname{C}(T, w)
  \label{cmax}
\end{equation}

where, $T$ is the event trace, $arg$ is an abbreviation for argument, $C$ is the count function, $w$ is the window size, and $W_\delta$ is the window size threshold.

\begin{figure}[t]
	\centering           
	\includegraphics[trim = 15mm 2mm 11mm 4mm, clip, width=0.5\textwidth]{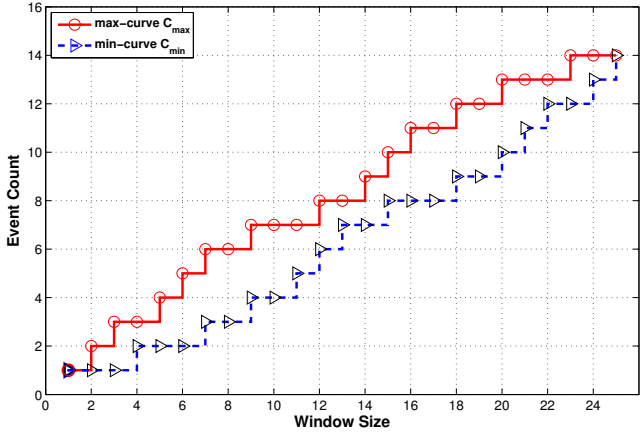}   
	\caption{Maximum and minimum inter-arrival curves for $e = B$.} 
	\label{IAC}        
\end{figure}

Consider the below event trace sample $T$. It contains 25 events of event types $e \in \{A, B, C, D, E\}$. The corresponding minimum inter-arrival curve and maximum inter-arrival curve for an event type $B$ with $W_\delta$ of size 25 are shown in Figure~\ref{IAC}.

\hfill \par

$T = \big\{BBEBCABEABDBBBEBCBAABBBEB\big\}$

\hfill \par

Initially, a set of pairs of $C_{min}$ and $C_{max}$ for every considered event $e$ per each training trace were built during the training phase. This phase basically requires a massive computation capabilities to compute the maximum and the minimum inter-arrival curves as well as a lot of memory space to store the huge sets of pairs. Therefore, the inter-arrival curves were calculated only for significance percentage of an event type $e$, feature events, within a trace to reduce the computation during this phase. During the second phase, the framework aggregated the similar minimum and maximum inter-arrival curves applied to the same event $e$. The aggregation was achieved by calculating a mean curves and a student’s t-test confidence intervals for $C_{min}$ and $C_{max}$ independently at each window size. In other words, the output of the second phase is a training model composed of six inter-arrival curves per each type of $e$ event. The main advantages of aggregation method are represented by its capability of reducing the required allocation memory for the training model and its robustness against outliers.

During classification phase, the classifier used the obtained features, $C_{min}$ and $C_{max}$, from the tested trace to evaluate their conformance to the corresponding features obtained from the set of known traces in the training model. Such an evaluation was performed in an automated manner by applying the Mann-Whitney U test \cite{macfarland2016mann}. The classification framework considers the behavior of an event to be anomalous when its $C_{min}$, $C_{max}$, or both are anomalous. If the minimum or the maximum inter-arrival curve failed the Mann-Whitney U test, the classifier performs further classification stage to quantify the deviation of the failed curve from the training model. It uses the tunable threshold for this purpose, the tunable threshold is referred to as $\mathlarger{\mathlarger{\mathlarger{\varsigma}}}_{th}$. The inter-arrival curve is declared to be normal if its deviation value is less than $\mathlarger{\mathlarger{\mathlarger{\varsigma}}}_{th}$, even that it failed the statistical Mann-Whitney U test. Otherwise, the inter-arrival curve is considered anomalous.

\section{Experimental Work}

The proposed framework has a dual detection that relies on the equipments readings with their thresholds and the inter-arrival curves. The detection system will make use of data and event traces, the data trace represents the equipments readings, such as the electrical and temperature, at timing durations for a given event trace. The system behavior is declared to be normal if the event trace and the data trace pass the inter-arrival curves test and the threshold-based test, respectively. Otherwise, the behavior is considered anomalous. Not that anomalies affecting events timing are detected by threshold-based framework. In addition, min and max inter-arrival curves are unable to detect anomalies which have a slight perturbation to the event trace as we mentioned previously. In this case, threshold-based anomaly detection framework takes care of detecting such abnormal behaviour.

We have used data of a turbine-generator operation from power generation plant. The purposed system uses threshold-based anomaly detection as well as event-based anomaly detection. The threshold is based on the operational limits and the average of a parameter. The dataset is evaluated using support vector machine, k-nearest neighbor, and C4.5 decision tree for threshold-based anomaly detection.

\subsection{Dataset and Feature Selection}

The dataset is comprised of 18 parameters related to temperature, pressure, electrical and safety indicators. The timing is very strict so it is not feasible to use all the parameters to identify anomalies. This is due to high incoming data rate and timely execution of all the processes in industrial applications.
Feature selection is carried on dataset to reduce the number of features to differentiate normal and anomalous industrial operation. Genetic and exhaustive search algorithms are used to identify important features in the dataset. Features that affect the power output of the turbine-generator are selected using feature selection algorithms. The feature space reduced from 18 to 5 important features only which greatly affect power production. The selected features are as follows:

\begin{itemize}
  \item Fuel Gas Flow (FGF): It is the amount of gas input to the combustion chamber. Its value varies as per the power requirement.
  \item Main Shaft Vibration (MSV): It is the shaft vibration intensity. It is very important parameter, any great deviation from normal operation can persistently damage the turbine-generator system.
  \item Gearbox Vibration (GBV): It is similar safety indicator as main shaft vibration.
  \item Exhaust Gas Temperature (EGT): It is temperature of turbine exhaust, which varies with the power output of the generator. It needs to be monitored because excessive exhaust temperature can also damage the turbine-generator system.
  \item Power: Several parameters such as compressor discharge pressure, lube oil supply temperature directly affect power output of a turbine-generator system. Therefore, power is also selected to identify abnormal operations in this process.
\end{itemize}

\begin{figure}[t]
	\centering           
	\includegraphics[trim = 15mm 4mm 11mm 7mm, clip, width=0.5\textwidth]{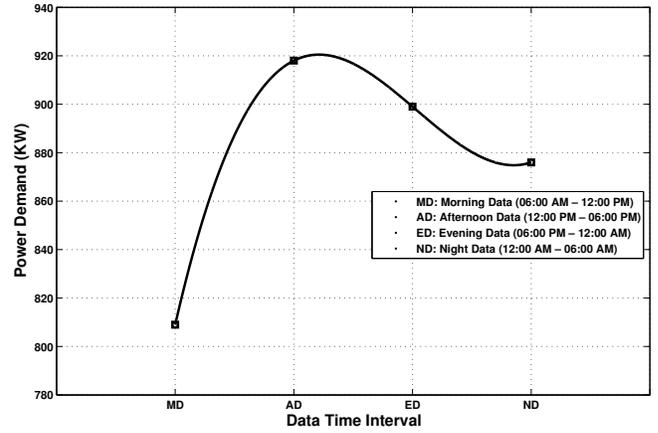} 
	\caption{Average power demand during different time intervals.} 
	\label{power}        
\end{figure}

Each parameter of turbine-generator system should fall in predefined range to ensure normal system operation. The power requirement varies during different times of the day. The static nature of these processes can be used to model the normal operation behavior of the system.
The baseline parameters are established using the historical data from a power generation plant. The dataset is divided based on different power demand patterns during different time intervals, Figure~\ref{power} shows the average power demand during different time intervals. The dataset divisions are as follows:

\begin{itemize}
  \item Morning Dataset (MD): Data from the morning time interval (06:00 AM - 12:00 PM).
  \item Afternoon Dataset (AD): Data from the afternoon time interval (12:00 PM - 06:00 PM).
  \item Evening Dataset (ED): Data from the evening time interval (06:00 PM - 12:00 AM).
  \item Night Dataset (ND): Data from the night time interval (12:00 AM - 06:00 AM).
\end{itemize}

\subsection{Testing Scenarios and Detection Sensitivity Degree ($S_\%$)}

To evaluate the performance of the proposed ADS, six testing scenarios are considered per dataset group. The difference between each testing scenario is based on the number of parameters considered for anomaly detection and the total number of parameters in the dataset. The testing scenarios are divided into two classes; reduced dataset class and full dataset class. The reduced dataset represents the datasets with the parameters that are selected from the original dataset using feature selection algorithms. The full dataset consists of all the parameters in the original dataset. This is done to evaluate the impact of the feature selection on the performance of the proposed ADS. In addition, three scenarios are selected for each class based on the sensitivity degree of detecting anomalies. These testing scenarios are classified as follows:

\begin{itemize}
  \item Least Sensitive ($S_\% = 20\%$): A dataset row is labeled as an attack if all the selected parameters in the reduced dataset are compromised. For the full dataset, the dataset row is considered anomalous when all parameters are compromised.
  \item Medium Sensitive ($S_\% = 60\%$): A dataset row is labeled anomalous when any three of the selected parameters in the reduced dataset or any three parameters in the full dataset are compromised.
  \item Highly Sensitive ($S_\% = 100\%$): A dataset row is labeled as an attack if any one of the parameters or the selected parameters in the full dataset or the reduced dataset is compromised. 
\end{itemize}

The maximum, minimum, trim means, and thresholds values are calculated for each parameter. Trim means are used instead of the normal averages to exclude the outliers that affect the average. This helps in protecting the normal profiles from an attacker attempting to manipulate them by sending periodic malicious values. It also protects the normal profiles from the values coming from the system faults.

The difference percentage between the operational limit and the trim mean is calculated using Equation~\ref{diff}. This percentage is used to calculate the tolerance region between the trim mean and the threshold. If the average value of a parameter is close to its operational limit, its tolerance region will be smaller as compared to a higher average value. For small tolerance region, small variation will result in an anomaly behavior.

\begin{equation}
  \bigtriangleup  = \frac{~|\psi - \mu|~}{\psi}
  \label{diff}
\end{equation}

where, $\bigtriangleup$ is the percentage difference of the operational limit and the trim mean, $\psi$ is the operational limit, and $\mu$ is the parameter trim mean.

\begin{figure*}[t!]
	\centering           
	\includegraphics[height = 7.5 cm, width=0.8\textwidth]{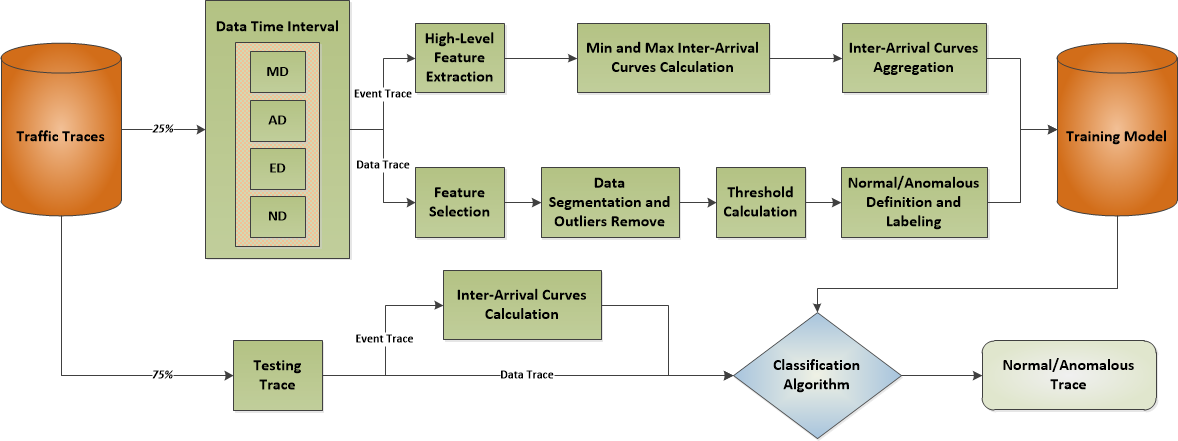}   
	\captionsetup{justification=centering}
	\caption{Proposed approach architecture.}  
	\label{Architecture}       
\end{figure*}

Equation~\ref{pth} is developed to determine the threshold for each parameter. It should be noted that the threshold is not static and incorporate the changing operational conditions. The parameter is considered to be compromised if its value exceeds the established threshold calculated using Equation~\ref{pth}.

\begin{equation}
  P_{th}  = \mu \cdot \bigtriangleup + \mu
  \label{pth}
\end{equation}

Sample rows of the used dataset with normal values for the selected parameters are shown in Table~\ref{tab_normal}, whereas Table~\ref{tab_anomalous} shows the anomalous values. These tables include the operational limit of each parameter, the computed thresholds, the normal values, and the anomalous values. The anomalous values are shown in bold italic font in Table~\ref{tab_anomalous}.

\begin{table}[b!]
	\renewcommand{\arraystretch}{1.2}
	\centering
	\captionsetup{justification=centering}
	\caption{Normal operation dataset.}
\begin{tabular}{c c c c c c}
\hline
\multirow{2}{*}{} & \multicolumn{5}{c}{Parameter}\\
\cline{2-6}
 & FGF & MSV & GBV & EGT & Power \\
\hline
 \hline
 $\psi$ & 500 & 45 & 5 & 560 & 1120 \\
\hline
 $P_{th}$ & 445 & 18.75 & 2.50 & 532 & 1032 \\
\hline
  Samples & 329 & 10.51 & 1.43 & 469 & 918 \\
  & 319 & 11.62 & 1.56 & 416 & 834 \\
  & 362 & 10.90 & 1.40 & 433 & 800 \\
  &  371 & 12.42 & 1.60 & 458 & 819 \\
\hline
\end{tabular}
	\label{tab_normal}
	\vspace*{0.75 cm}
	
	\renewcommand{\arraystretch}{1.2}
	\centering
	\captionsetup{justification=centering}
	\caption{Anomalous operation dataset.}
\begin{tabular}{c c c c c c}
\hline
\multirow{2}{*}{} & \multicolumn{5}{c}{Parameter}\\
\cline{2-6}
 & FGF & MSV & GBV & EGT & Power \\
\hline
 \hline
 $\psi$ & 500 & 45 & 5 & 560 & 1120 \\
\hline
 $P_{th}$ & 445 & 18.75 & 2.50 & 532 & 1032 \\
\hline
  Samples & \textbf{\textit{484}} & 12.636 & 1.365 & 470 & 884 \\
  & \textbf{\textit{474}} & 10.998 & 1.425 & \textbf{\textit{547}} & \textbf{\textit{1078}} \\
  & \textbf{\textit{447}} & \textbf{\textit{23.51}} & \textbf{\textit{4.56}} & \textbf{\textit{557}} & \textbf{\textit{1103}} \\
\hline
\end{tabular}
	\label{tab_anomalous}
	
\end{table}

For the event-based anomaly detection, the minimum and the maximum inter-arrival curves are calculated and tested for the events related to the considered parameters in the event trace. The number of the considered events depends on the value of $S_\%$ when using the full dataset class. The most significant 20\% of the events are considered by the inter-arrival curves anomaly detection when testing the least sensitive scenario. On the other hand, 60\% of the significant events are tested by the event-based anomaly detection during the medium sensitive scenario. However, the inter-arrival curves for all events are generated and tested for the highly sensitive scenario. The events related to the 5 important selected features are considered by the inter-arrival curves anomaly detection when using the reduced dataset class. Subsequently, the event trace will be considered anomalous when all events, any three events, and any event fail the inter-arrival curves test for the least sensitive, medium sensitive, and highly sensitive scenarios, respectively.

\subsection{Performance Measures}

The performance evaluation of the proposed system is done using three important factors. The key performance indicators include the Attack Detection Rate (ADR), the False Positive Rate (FPR), and the System Accuracy (SA). The ADR, FPR, and SA are calculated using Equation~\ref{ADR}, Equation~\ref{FPR}, and Equation~\ref{SA}, respectively \cite{shrivastava2011effective}.

\begin{equation}
  ADR  = \frac{Total~No.~of~attacks}{Total~No.~of~detected~attacks}\times100\%
  \label{ADR}
\end{equation}

\begin{equation}
  FPR  = \frac{Total~No.~of~misclassified~Proc}{Total~No.~of~normal~Proc}\times100\%
  \label{FPR}
\end{equation}

\begin{equation}
  SA  = \frac{Total~No.~of~right~classified~Proc}{Total~No.~of~Proc}\times100\%
  \label{SA}
\end{equation}

\vspace*{0.14 cm}

The proposed anomaly detection system targets the classification problem through a framework whose architecture is shown in Figure~\ref{Architecture}. Using the given training set of data and event traces, which also known as training traces, the output of the framework is a decision whether a test trace is conforming to that training data; normal trace, or anomalous trace. The framework uses the threshold-based anomaly detection techniques as well as the inter-arrival curves for extracting the high-level features from the data and event traces, respectively. Subsequently, it builds a training model using these features to reason about the corresponding features extracted from the test trace.

\section{Results and Analysis}

The dataset comprised of a one month of operational data from the power generation process. The training of the algorithms is done over 25\% of a randomly selected data. The smaller training set data is used because of the fewer number of attack types, similar approach is used in \cite{goldenberg2013accurate} where minimal training is done for the IDS. The data/event traces used in the testing phase are different from the traces used for training the algorithms. Each instance of the trained dataset is labelled as normal or anomalous. The labeling of the training dataset is done according to the following criteria, the normal activity is represented by +1 while the anomalous is represented by -1. The total of 18 features are used to train the classifier algorithm in all features based anomaly detection, whereas 5 selected features by the feature selection algorithms are used to train the algorithms in the reduced feature dataset based anomaly detection.

The implementation of the algorithms is done through CUDA C/C++ programming language on HPC K-20 GPU cluster \cite{dongarra2013toward} to achieve the parallel computations which reduce the computation time.  The cluster is comprised of 32 IBM System x3650 servers, dual Xeon E5 10-core 2.8 gigahertz processors, NVIDIA K20 GPUs containing 2,496 cores each on 12 of these nodes, with 64 gigabytes of memory.

The experiments tested the attack detection rate, the false positive rate, and the accuracy factors among full dataset and reduced feature input to the classifier algorithm. The results of the experiments are presented for each type of the attack sensitivity degree (i.e., least sensitive, medium sensitive, and highly sensitive). The experimental results of each time interval group are presented for each type of classification algorithm along with their averages.

\subsection{Support Vector Machine (SVM)}

The experimental results of the proposed approach with $S_\%$ as $100\%$ are summarized in Table~\ref{svm_s100}. The achieved average accuracy with the reduced dataset is 87.8\% whereas it is 71.3\% for full dataset based approach. Performance degradation was observed for MD group as the detection accuracy dropped to 80.6\%. The minimum accuracy with full dataset based approach is 57.8\% for ND group. This low detection accuracy is observed due to the similarity between the normal and the anomalous data. The average attack detection rate for reduced data set based approach is 85\%. The full dataset based approach achieved the maximum attack detection rate of 55.6\% with the average detection rate of 46.7\%. The average false positive rate with reduced dataset input is 11.3\% whereas with the full dataset it is 20.6\%. ND group showed the maximum false positive rate of 41.5\% for the full dataset.

\begin{table}[t!]
	\renewcommand{\arraystretch}{1.2}
	\centering
	\captionsetup{justification=centering}
	\caption{SVM results for $S_\%=100\%$.}
	\resizebox{0.95\columnwidth}{!}{
\begin{tabular}{lSSSSSSSS}
    \toprule
    \multirow{2}{*}{} &
      \multicolumn{4}{c}{Full Dataset} &
      \multicolumn{4}{c}{Reduced Dataset} \\
      & {MD} & {AD} & {ED} & {ND} & {MD} & {AD} & {ED} & {ND} \\
      \hline
      \midrule
    ADR & 55.6 & 37.8 & 37.8 & 55.6 & 71.1 & 84.4 & 93.3 & 91.1 \\
    FPR & 23.7 & 11.9 & 5.2 & 41.5 & 6.7 & 20.7 & 6.7 & 11.1 \\
    SA & 71.1 & 75.6 & 80.6 & 57.8 & 87.8 & 80.6 & 93.3 & 89.4 \\
    \bottomrule
  \end{tabular}
  }
	\label{svm_s100}
\end{table}

The experimental results improved greatly by increasing the number of parameters, $S_\% = 60\%$, for anomaly detection as shown in Table~\ref{svm_s60}. The ideal accuracy of 100\% is achieved for AD and ED groups with reduced dataset. The average accuracy is also very close to ideal. Another benefit is ideal false positive rate of zero across all groups. The attack detection rate also improved from 85\% to around 98\%. The improved results are due to improved discrimination criteria between normal and anomalous data. Similar results are observed for full dataset approach, its average accuracy increased to 92.1\% but its attack detection rate still very low. The maximum achieved attack detection rate is 77.8\% with average value of 68.9\%. Similar to reduce dataset approach, the false positive rate is very close to the ideal condition. In this approach, increasing the parameter increase the accuracy and the false positive rate but the attack detection rate still not appropriate for industrial control systems. The poor attack detection rate is due to the poor discriminating criteria which is due to weak support vectors in hyperplane.

\begin{table}[b!]
	\renewcommand{\arraystretch}{1.2}
	\centering
	\captionsetup{justification=centering}
	\caption{SVM results for $S_\%=60\%$.}
	\resizebox{0.95\columnwidth}{!}{
\begin{tabular}{lSSSSSSSS}
    \toprule
    \multirow{2}{*}{} &
      \multicolumn{4}{c}{Full Dataset} &
      \multicolumn{4}{c}{Reduced Dataset} \\
      & {MD} & {AD} & {ED} & {ND} & {MD} & {AD} & {ED} & {ND} \\
      \hline
      \midrule
    ADR & 57.8 & 77.8 & 75.6 & 64.4 & 95.6 & 100.0 & 100.0 & 95.6 \\
    FPR & 0.0 & 0.0 & 0.7 & 0.0 & 0.0 & 0.0 & 0.0 & 0.0 \\
    SA & 89.4 & 94.4 & 93.3 & 91.1 & 98.9 & 100.0 & 100.0 & 98.9 \\
    \bottomrule
  \end{tabular}
  }
	\label{svm_s60}
	\vspace*{0.45 cm}
\end{table}

\begin{table}[b!]
	\renewcommand{\arraystretch}{1.2}
	\centering
	\captionsetup{justification=centering}
	\caption{SVM results for $S_\%=20\%$.}
	\resizebox{0.95\columnwidth}{!}{
\begin{tabular}{lSSSSSSSS}
    \toprule
    \multirow{2}{*}{} &
      \multicolumn{4}{c}{Full Dataset} &
      \multicolumn{4}{c}{Reduced Dataset} \\
      & {MD} & {AD} & {ED} & {ND} & {MD} & {AD} & {ED} & {ND} \\
      \hline
      \midrule
    ADR & 84.4 & 75.6 & 66.7 & 64.4 & 100.0 & 100.0 & 100.0 & 100.0 \\
    FPR & 0.0 & 0.0 & 0.0 & 0.0 & 0.0 & 0.0 & 0.0 & 0.0 \\
    SA & 96.1 & 93.9 & 91.7 & 91.1 & 100.0 & 100.0 & 100.0 & 100.0 \\
    \bottomrule
  \end{tabular}
  }
	\label{svm_s20}
\end{table}

The ideal attack detection rate, accuracy, and false positive rate are achieved for the least sensitive detection system across all groups with reduced dataset as shown in Table~\ref{svm_s20}. The difference between the normal and the anomalous classes is very high because there is no similarity between them since all the parameters are either anomalous or normal in a particular row. This establishes optimal hyperplane between the normal and the anomalous class. The results with the full dataset also shows improvement. The average accuracy is improved to 96.7\% with ideal false positive rate across all groups. However, this approach still suffers in the attack detection rate. Despite improvement in some groups, the average attack detection rate only improves slightly. The maximum attack detection rate is about 84.4\% for MD group but the average attack detection rate is around 73\%.

\begin{figure}[t]
	\centering           
	\includegraphics[trim = 15mm 4mm 11mm 7mm, clip, width=0.5\textwidth]{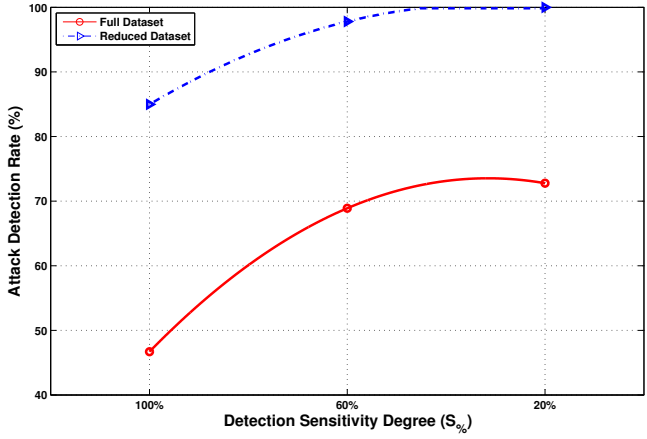} 
	\caption{Average attack detection rate for SVM.} 
	\label{svmadr}        
\end{figure}

\begin{figure}[t]
	\centering           
	\includegraphics[trim = 15mm 4mm 11mm 7mm, clip, width=0.5\textwidth]{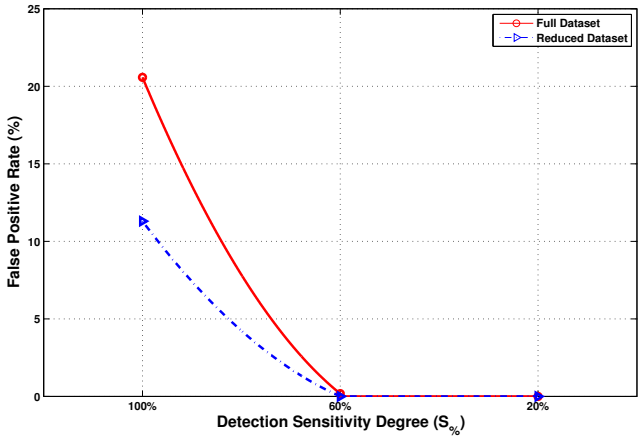} 
	\caption{Average false positive rate for SVM.} 
	\label{svmfpr}        
\end{figure}

\begin{figure}[t]
	\centering           
	\includegraphics[trim = 15mm 4mm 11mm 7mm, clip, width=0.5\textwidth]{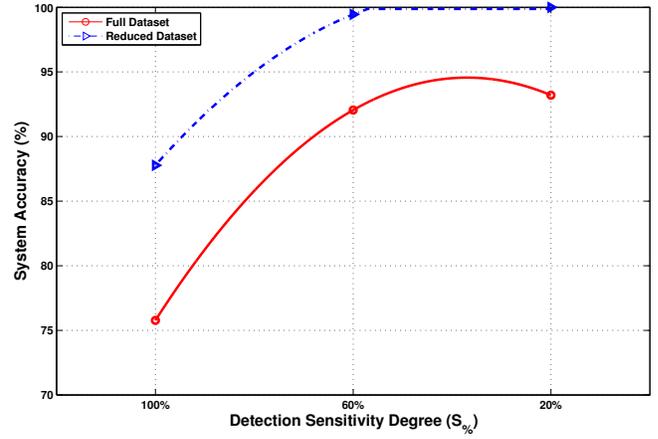} 
	\caption{Average accuracy for SVM.} 
	\label{svmsa}        
\end{figure}

As a summary, the results show that increasing the number parameters (i.e., reducing the sensitivity degree) for anomaly detection improves the performance of the system when considering SVM as a classification algorithm. The reduced feature set outperform the full dataset based approach in all the testing scenarios. The result of 5 parameters based approach are ideal but this approach is also least sensitive to the attacks. The one parameter based approach is highly sensitive to attacks but the suffers from low accuracy and attack detection rate. Its performance can also be degraded due to the noise which is quite common in industrial control system. The 3 parameters based provide balance between the sensitivity to attack and the performance characteristics. It has ideal false positive rate, high accuracy, and attack detection rate. The comparison of attack detection rate, false positive rate, and accuracy between the reduced dataset and the full dataset based approach are shown in Figure~\ref{svmadr}, Figure~\ref{svmfpr}, and Figure~\ref{svmsa}, respectively.

\subsection{K-Nearest Neighbor (k-NN)}

The results of each group considering one parameter based approach are shown in Table~\ref{knn_s100}. k-NN does not perform well for either reduced dataset or full dataset based approach. The attack detection rate is very low in both the cases. In reduced dataset, the maximum attack detection rate is approximately 38\%, which belongs to ED group. The achieved average detection rate is 29\%. The full dataset based approach follows the same trend with average attack rate as 17\% only. Beside low attack detection rate, this approach also has considerable high false positive rate for full dataset based approach. The average false positive rate is 15.5\% for reduced dataset based anomaly detection whereas it reaches to 28\% when full dataset is used.

\begin{table}[b!]
	\renewcommand{\arraystretch}{1.2}
	\centering
	\captionsetup{justification=centering}
	\caption{K-NN results for $S_\%=100\%$.}
	\resizebox{0.95\columnwidth}{!}{
\begin{tabular}{lSSSSSSSS}
    \toprule
    \multirow{2}{*}{} &
      \multicolumn{4}{c}{Full Dataset} &
      \multicolumn{4}{c}{Reduced Dataset} \\
      & {MD} & {AD} & {ED} & {ND} & {MD} & {AD} & {ED} & {ND} \\
      \hline
      \midrule
    ADR & 20.0 & 13.3 & 13.3 & 20.0 & 24.4 & 33.3 & 37.8 & 20.0  \\
    FPR & 32.6 & 23.0 & 34.1 & 23.0 & 14.1 & 18.5 & 11.9 & 17.8  \\
    SA & 55.6 & 61.1 & 52.8 & 62.8 & 70.6 & 69.4 & 75.6 & 66.7  \\
    \bottomrule
  \end{tabular}
  }
	\label{knn_s100}
	\vspace*{0.45 cm}
\end{table}

\begin{table}[b!]
	\renewcommand{\arraystretch}{1.2}
	\centering
	\captionsetup{justification=centering}
	\caption{K-NN results for $S_\%=60\%$.}
	\resizebox{0.95\columnwidth}{!}{
\begin{tabular}{lSSSSSSSS}
    \toprule
    \multirow{2}{*}{} &
      \multicolumn{4}{c}{Full Dataset} &
      \multicolumn{4}{c}{Reduced Dataset} \\
      & {MD} & {AD} & {ED} & {ND} & {MD} & {AD} & {ED} & {ND} \\
      \hline
      \midrule
    ADR & 26.7 & 28.9 & 24.4 & 26.7 & 75.6 & 68.9 & 77.8 & 75.6   \\
    FPR & 20.0 & 22.2 & 25.2 & 16.3 & 16.3 & 3.7 & 8.1 & 11.1 \\
    SA & 66.7 & 65.6 & 62.2 & 69.4 & 81.7 & 89.4 & 88.3 & 85.6  \\
    \bottomrule
  \end{tabular}
  }
	\label{knn_s60}
\end{table}

The results are considerably improved for reduced features based approach when the number of parameters for anomaly detection increased to three as shown in Table~\ref{knn_s60}. The maximum attack detection rate increased to 77.8\% whereas the average is improved to 74.4\%. However, such attack detection rate does not meet the safety requirement of real industrial control systems. The full dataset based approach still suffers from low attack detection rate. The average attack detection rate is 26.67\% with an improvement of about 10\%. Similarly, the accuracy and the false positive rate also improve in this approach. The minimum achieved false positive rate for reduced feature set is 3.7\% but average is 9.8\%. The average accuracy is improved to 86.25\%. In case of full dataset based approach, the average accuracy increased to 66\% with an improvement of 8\%.

\begin{figure}[t]
	\centering           
	\includegraphics[trim = 15mm 4mm 11mm 7mm, clip, width=0.5\textwidth]{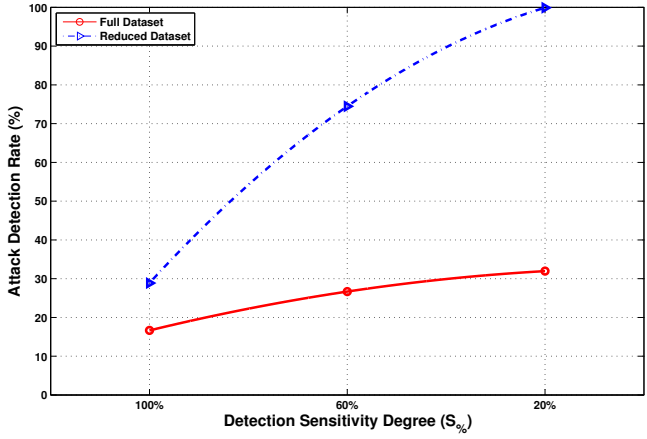} 
	\caption{Average attack detection rate for k-NN.} 
	\label{knnadr}        
\end{figure} 

\begin{figure}[t]
	\centering           
	\includegraphics[trim = 15mm 4mm 11mm 7mm, clip, width=0.5\textwidth]{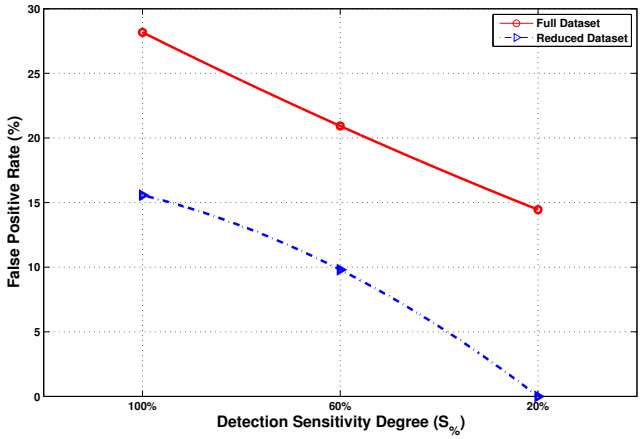} 
	\caption{Average false positive rate for k-NN.} 
	\label{knnfpr}        
\end{figure}

\begin{table}[b]
	\renewcommand{\arraystretch}{1.2}
	\centering
	\captionsetup{justification=centering}
	\caption{K-NN results for $S_\%=20\%$.}
	\resizebox{0.95\columnwidth}{!}{
\begin{tabular}{lSSSSSSSS}
    \toprule
    \multirow{2}{*}{} &
      \multicolumn{4}{c}{Full Dataset} &
      \multicolumn{4}{c}{Reduced Dataset} \\
      & {MD} & {AD} & {ED} & {ND} & {MD} & {AD} & {ED} & {ND} \\
      \hline
      \midrule
    ADR & 31.1 & 31.1 & 26.7 & 28.9 & 100.0 & 100.0 & 100.0 & 100.0  \\
    FPR & 12.6 & 7.4 & 23.7 & 14.1 & 0.0 & 0.0 & 0.0 & 0.0 \\
    SA & 73.3 & 77.2 & 63.9 & 71.7 & 100.0 & 100.0 & 100.0 & 100.0 \\
    \bottomrule
  \end{tabular}
  }
	\label{knn_s20}
\end{table}

\begin{figure}[t]
	\centering           
	\includegraphics[trim = 15mm 4mm 11mm 7mm, clip, width=0.5\textwidth]{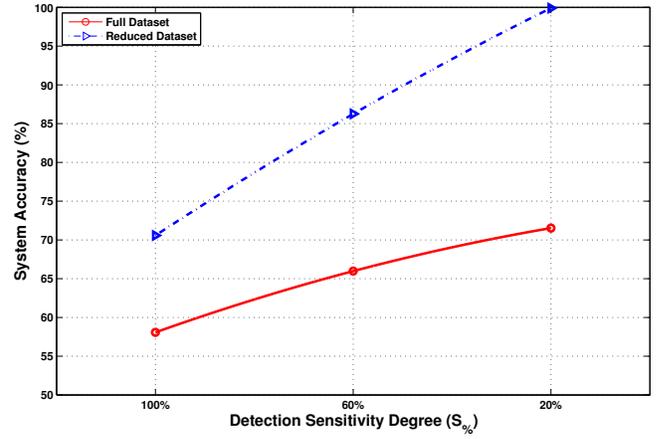} 
	\caption{Average accuracy for k-NN.} 
	\label{knnsa}        
\end{figure}

Similar to SVM, the proposed approach with feature reduction produce ideal accuracy, false positive rate, and attack detection rate for $S_\% = 20\%$ as shown in Table~\ref{knn_s20}. As all features are either in normal or anomalous region, this results in high distance between the normal and the anomalous classes. The full dataset based approach does not show considerable improvement in the attack detection rate. The maximum attack detection rate only improved by 2\% whereas the average attack detection rate is 29.4\%. In this case, the attack trace consists of 18 parameters with 5 of them in the anomalous region. The distance calculation between the testing sample and the training data considers 13 normal features, this results in choosing incorrect nearest neighbor. This is evident due to decrease in the false positive rate. The average false positive rate is 14.4\%, the average accuracy is improved to 71.5\%.

As a summary, k-NN results follows the same trend of SVM and its results improved by increasing the number of parameters. The average attack detection rate, false positive rate, and accuracy are shown in Figure~\ref{knnadr}, Figure~\ref{knnfpr}, and Figure~\ref{knnsa}, respectively. The one parameter based approach does not produce acceptable result neither from reduced dataset nor the full dataset approach. The three parameters based approach shows improvement in the results. However, the attack detection rate does not show much improvement with full dataset. The attack detection rate for this approach continues to suffer even with five parameters based approach. The attack detection rate for the reduced dataset based approach shows high improvement but may be not appropriate for industrial application since the average attack detection rate is 74.4\%. The ideal results are achieved with reduced dataset and five parameters approach.

\subsection{C4.5 Decision Tree}

The experimental results for the proposed approach with one parameter based anomaly detection are summarized in Table~\ref{ctree_s100}. The achieved average accuracy with reduced feature set is 92.4\%. The average accuracy suffers due to the low accuracy of 88.3\% in AD group. This implies that wrong discriminating features are selected in the classification criteria. In case of full dataset, the achieved average accuracy is 72.6\%. Performance degradation is observed
for MD group as the detection accuracy dropped to 65\%. Moreover, the average attack detection rate for the reduced feature based approach is 71.7\%. The full dataset based approach achieved a maximum attack detection rate of 20.0\% with the average detection rate of 15.0\%. On the other hand, the false positive rate is extremely low in all groups for reduced feature set based approach. This implies that the accuracy of this approach is affected by the number of false positives. The average false positive rate is 0.7\%. In case of the full dataset based approach, the average false positive rate is 8.1\%.

\begin{table}[b!]
	\renewcommand{\arraystretch}{1.2}
	\centering
	\captionsetup{justification=centering}
	\caption{C4.5 results for $S_\%=100\%$.}
	\resizebox{0.95\columnwidth}{!}{
\begin{tabular}{lSSSSSSSS}
    \toprule
    \multirow{2}{*}{} &
      \multicolumn{4}{c}{Full Dataset} &
      \multicolumn{4}{c}{Reduced Dataset} \\
      & {MD} & {AD} & {ED} & {ND} & {MD} & {AD} & {ED} & {ND} \\
      \hline
      \midrule
    ADR & 15.6 & 13.3 & 11.1 & 20.0 & 80.0 & 53.3 & 73.3 & 80.0   \\
    FPR & 18.5 & 0.0 & 2.2 & 11.9 & 1.5 & 0.0 & 0.0 & 1.5   \\
    SA & 65.0 & 78.3 & 76.1 & 71.1 & 93.9 & 88.3 & 93.3 & 93.9   \\
    \bottomrule
  \end{tabular}
  }
	\label{ctree_s100}
	\vspace*{0.45 cm}
\end{table}

\begin{table}[b!]
	\renewcommand{\arraystretch}{1.2}
	\centering
	\captionsetup{justification=centering}
	\caption{C4.5 results for $S_\%=60\%$.}
	\resizebox{0.95\columnwidth}{!}{
\begin{tabular}{lSSSSSSSS}
    \toprule
    \multirow{2}{*}{} &
      \multicolumn{4}{c}{Full Dataset} &
      \multicolumn{4}{c}{Reduced Dataset} \\
      & {MD} & {AD} & {ED} & {ND} & {MD} & {AD} & {ED} & {ND} \\
      \hline
      \midrule
    ADR & 55.6 & 44.4 & 46.7 & 60.0 & 100.0 & 86.7 & 91.1 & 88.9   \\
    FPR & 0.0 & 8.9 & 5.9 & 0.7 & 4.4 & 0.0 & 0.0 & 0.0  \\
    SA & 88.9 & 79.4 & 82.2 & 89.4 & 96.7 & 96.7 & 97.8 & 97.2 \\
    \bottomrule
  \end{tabular}
  }
	\label{ctree_s60}
\end{table}

The results are considerably improved by increasing the number of parameter for anomaly detection to three as shown in Table~\ref{ctree_s60}. The average attack detection rate of the full dataset based approach is improved to 51.7\%. However, it still not suitable for practical industrial control security solution. The improvement in the attack detection rate is reflected to the accuracy. The accuracy for full dataset based approach is improved to 85.0\%. The false positive rate dropped to 3.9\%. Similarly, the results for reduced feature set based approach also improved. The average attack detection rate is 91.7\%. The accuracy is improved by 4.7\% to be 97.1\%. The false positive rate is slightly increased to 1.1\%.

\begin{figure}[t]
	\centering           
	\includegraphics[trim = 15mm 4mm 11mm 7mm, clip, width=0.5\textwidth]{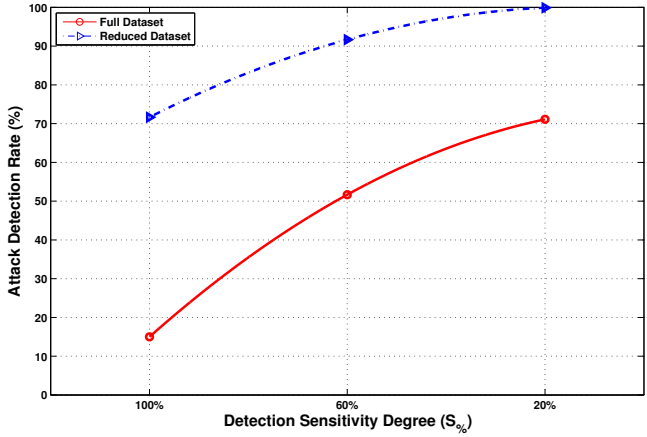} 
	\caption{Average attack detection rate for C4.5.} 
	\label{ctreeadr}        
\end{figure} 

\begin{figure}[t]
	\centering           
	\includegraphics[trim = 15mm 4mm 11mm 7mm, clip, width=0.5\textwidth]{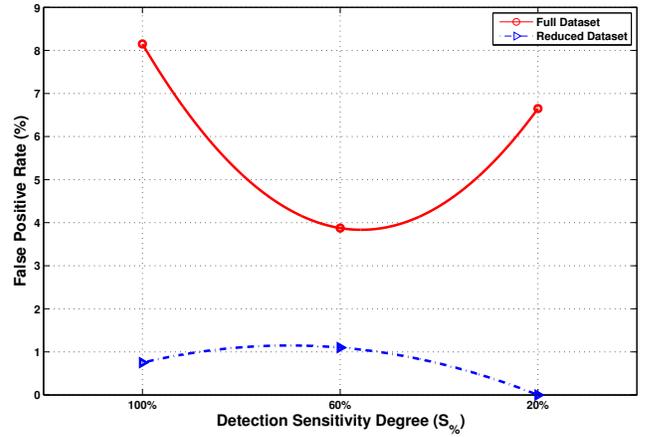} 
	\caption{Average false positive rate for C4.5.} 
	\label{ctreefpr}        
\end{figure}

\begin{table}[b]
	\renewcommand{\arraystretch}{1.2}
	\centering
	\captionsetup{justification=centering}
	\caption{C4.5 results for $S_\%=20\%$.}
	\resizebox{0.95\columnwidth}{!}{
\begin{tabular}{lSSSSSSSS}
    \toprule
    \multirow{2}{*}{} &
      \multicolumn{4}{c}{Full Dataset} &
      \multicolumn{4}{c}{Reduced Dataset} \\
      & {MD} & {AD} & {ED} & {ND} & {MD} & {AD} & {ED} & {ND} \\
      \hline
      \midrule
    ADR & 73.3 & 57.8 & 68.9 & 84.4 & 100.0 & 100.0 & 100.0 & 100.0  \\
    FPR & 9.6 & 0.0 & 3.7 & 13.3 & 0.0 & 0.0 & 0.0 & 0.0 \\
    SA & 86.1 & 89.4 & 89.4 & 86.1 & 100.0 & 100.0 & 100.0 & 100.0\\
    \bottomrule
  \end{tabular}
  }
	\label{ctree_s20}
\end{table}

\begin{figure}[t]
	\centering           
	\includegraphics[trim = 15mm 4mm 11mm 7mm, clip, width=0.5\textwidth]{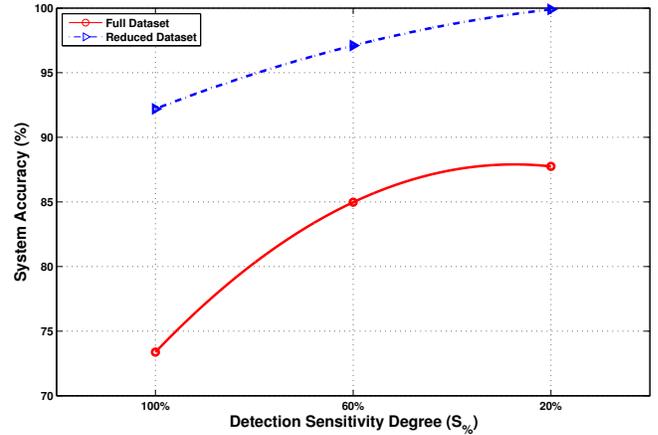} 
	\caption{Average accuracy for C4.5.} 
	\label{ctreesa}        
\end{figure}

The ideal results with reduced dataset are achieved with five parameter based approach as shown in Table~\ref{ctree_s20}, which is similar to SVM and k-NN results. The analysis of the decision tree shows that the decision is based on the most discriminating feature. This results in a small tree size with a binary decision logic. The average attack detection rate improved to 71.1\% with highest attack detection rate in ND group. The achieved average accuracy is also improved to 87.8\%. However, the false positive rate is increased to 6.7\% when compare to three parameters based approach. The average attack detection rate, false positive rate, and accuracy are shown in Figure~\ref{ctreeadr}, Figure~\ref{ctreefpr}, and Figure~\ref{ctreesa}, respectively. The reduced features set based approach outperforms the full dataset based approach in all test cases. The ideal results are achieved with five parameters based approach but the three parameters based approach gives the balance between the performance and the sensitivity to attacks.

\section{Conclusion}

In this paper, we proposed an Anomaly Detection System (ADS) for industrial control system and evaluate its performance. The novel component of this approach is the utilization of the physical parameters of the industrial process to detect anomalies. The proposed ADS uses the threshold-based anomaly detection technique as well as the inter-arrival curves for extracting the high-level features from the data and event traces, respectively. It builds a training model using these features to reason about the corresponding features extracted from the test trace. The system behavior is declared to be normal if the event trace and the data trace pass the inter-arrival curves test and the threshold-based test, respectively. Otherwise, the behavior is considered anomalous. 

The performance of the proposed solution enhanced by using feature selection method to reduce the feature set for anomaly detection. Support Vector Machine (SVM), k-Nearest Neighbor (k-NN), and C4.5 decision tree algorithm are used to train and test the proposed approach with the min and max inter-arrival curves. SVM and C4.5 produced accurate results even for high and medium sensitivity attacks. As compared to this, k-NN was unable to produce good results for low and medium sensitivity attacks test cases. The SVM based approach achieved ideal results with three parameters based approach, while k-NN achieved the same results with five parameters based approach. It also has low false positive rate and high detection rate as compare to k-NN. The study on the industrial control system and the proposed ADS results show that information about physical states of a process can be utilized to develop a security solutions for industrial control system. 

The future work involves the testing of the proposed ADS in different industrial control systems such as Oil and Gas industries. Future work involves the correlation of multiple anomaly detection systems to increase the accuracy of the system and testing the proposed approach on multiple industrial processes.

\section{Acknowledgment}

The authors would like to thank King Fahd University of Petroleum and Minerals (KFUPM) for supporting this research and providing the computing facilities.

\appendices



\ifCLASSOPTIONcaptionsoff
  \newpage
\fi


%

\bibliography{main}

\begin{thebibliography}{10}
\providecommand{\url}[1]{#1}
\csname url@samestyle\endcsname
\providecommand{\newblock}{\relax}
\providecommand{\bibinfo}[2]{#2}
\providecommand{\BIBentrySTDinterwordspacing}{\spaceskip=0pt\relax}
\providecommand{\BIBentryALTinterwordstretchfactor}{4}
\providecommand{\BIBentryALTinterwordspacing}{\spaceskip=\fontdimen2\font plus
\BIBentryALTinterwordstretchfactor\fontdimen3\font minus
  \fontdimen4\font\relax}
\providecommand{\BIBforeignlanguage}[2]{{%
\expandafter\ifx\csname l@#1\endcsname\relax
\typeout{** WARNING: IEEEtran.bst: No hyphenation pattern has been}%
\typeout{** loaded for the language `#1'. Using the pattern for}%
\typeout{** the default language instead.}%
\else
\language=\csname l@#1\endcsname
\fi
#2}}
\providecommand{\BIBdecl}{\relax}
\BIBdecl

\bibitem{teixeira2011cyber}
A.~Teixeira, G.~D{\'a}n, H.~Sandberg, and K.~H. Johansson, ``A cyber security
  study of a scada energy management system: Stealthy deception attacks on the
  state estimator,'' \emph{IFAC Proceedings Volumes}, vol.~44, no.~1, pp.
  11\,271--11\,277, 2011.

\bibitem{sommestad2010scada}
T.~Sommestad, G.~N. Ericsson, and J.~Nordlander, ``Scada system cyber
  security—a comparison of standards,'' in \emph{IEEE PES General
  Meeting}.\hskip 1em plus 0.5em minus 0.4em\relax IEEE, 2010, pp. 1--8.

\bibitem{galloway2013introduction}
B.~Galloway and G.~P. Hancke, ``Introduction to industrial control networks,''
  \emph{IEEE Communications surveys \& tutorials}, vol.~15, no.~2, pp.
  860--880, 2013.

\bibitem{bhuyan2014network}
M.~H. Bhuyan, D.~K. Bhattacharyya, and J.~K. Kalita, ``Network anomaly
  detection: methods, systems and tools,'' \emph{IEEE Communications Surveys \&
  Tutorials}, vol.~16, no.~1, pp. 303--336, 2014.

\bibitem{cagalaban2010improving}
G.~Cagalaban, T.~Kim, and S.~Kim, ``Improving scada control systems security
  with software vulnerability analysis,'' in \emph{Proceedings of the 12th
  WSEAS international conference on Automatic control, modelling \&
  simulation}.\hskip 1em plus 0.5em minus 0.4em\relax World Scientific and
  Engineering Academy and Society (WSEAS), 2010, pp. 409--414.

\bibitem{fovino2010taxonomy}
I.~N. Fovino, A.~Coletta, and M.~Masera, ``Taxonomy of security solutions for
  the scada sector,'' \emph{Project ESCORTS Deliverable}, vol.~2, 2010.

\bibitem{nicholson2012scada}
A.~Nicholson, S.~Webber, S.~Dyer, T.~Patel, and H.~Janicke, ``Scada security in
  the light of cyber-warfare,'' \emph{Computers \& Security}, vol.~31, no.~4,
  pp. 418--436, 2012.

\bibitem{zhu2011taxonomy}
B.~Zhu, A.~Joseph, and S.~Sastry, ``A taxonomy of cyber attacks on scada
  systems,'' in \emph{Internet of things (iThings/CPSCom), 2011 international
  conference on and 4th international conference on cyber, physical and social
  computing}.\hskip 1em plus 0.5em minus 0.4em\relax IEEE, 2011, pp. 380--388.

\bibitem{rautmare2011scada}
S.~Rautmare, ``Scada system security: Challenges and recommendations,'' in
  \emph{2011 Annual IEEE India Conference}.\hskip 1em plus 0.5em minus
  0.4em\relax IEEE, 2011, pp. 1--4.

\bibitem{johnson2010survey}
R.~E. Johnson, ``Survey of scada security challenges and potential attack
  vectors,'' in \emph{Internet Technology and Secured Transactions (ICITST),
  2010 International Conference for}.\hskip 1em plus 0.5em minus 0.4em\relax
  IEEE, 2010, pp. 1--5.

\bibitem{barbosa2010intrusion}
R.~R.~R. Barbosa and A.~Pras, ``Intrusion detection in scada networks,'' in
  \emph{IFIP International Conference on Autonomous Infrastructure, Management
  and Security}.\hskip 1em plus 0.5em minus 0.4em\relax Springer, 2010, pp.
  163--166.

\bibitem{barbosa2012difficulties}
R.~R. Barbosa, R.~Sadre, and A.~Pras, ``Difficulties in modeling scada traffic:
  a comparative analysis,'' in \emph{International Conference on Passive and
  Active Network Measurement}.\hskip 1em plus 0.5em minus 0.4em\relax Springer,
  2012, pp. 126--135.

\bibitem{gao2010scada}
W.~Gao, T.~Morris, B.~Reaves, and D.~Richey, ``On scada control system command
  and response injection and intrusion detection,'' in \emph{eCrime Researchers
  Summit (eCrime), 2010}.\hskip 1em plus 0.5em minus 0.4em\relax IEEE, 2010,
  pp. 1--9.

\bibitem{cardenas2011attacks}
A.~A. C{\'a}rdenas, S.~Amin, Z.-S. Lin, Y.-L. Huang, C.-Y. Huang, and
  S.~Sastry, ``Attacks against process control systems: risk assessment,
  detection, and response,'' in \emph{Proceedings of the 6th ACM symposium on
  information, computer and communications security}.\hskip 1em plus 0.5em
  minus 0.4em\relax ACM, 2011, pp. 355--366.

\bibitem{garitano2011review}
I.~Garitano, R.~Uribeetxeberria, and U.~Zurutuza, ``A review of scada anomaly
  detection systems,'' in \emph{Soft Computing Models in Industrial and
  Environmental Applications, 6th International Conference SOCO 2011}.\hskip
  1em plus 0.5em minus 0.4em\relax Springer, 2011, pp. 357--366.

\bibitem{sridhar2012cyber}
S.~Sridhar, A.~Hahn, and M.~Govindarasu, ``Cyber--physical system security for
  the electric power grid,'' \emph{Proceedings of the IEEE}, vol. 100, no.~1,
  pp. 210--224, 2012.

\bibitem{chikalov2013rough}
I.~Chikalov, V.~Lozin, I.~Lozina, M.~Moshkov, H.~S. Nguyen, A.~Skowron, and
  B.~Zielosko, ``Rough sets,'' in \emph{Three Approaches to Data
  Analysis}.\hskip 1em plus 0.5em minus 0.4em\relax Springer, 2013, pp.
  69--135.

\bibitem{pisicua2010rough}
I.~Pisic{\u{a}} and P.~Postolache, ``Rough set theory and its applications in
  electrical power engineering. a survey,'' \emph{UPB Sci. Bull., Series C},
  vol. 729, 2010.

\bibitem{zhu2010scada}
B.~Zhu and S.~Sastry, ``Scada-specific intrusion detection/prevention systems:
  a survey and taxonomy,'' in \emph{Proceedings of the 1st Workshop on Secure
  Control Systems (SCS)}, 2010.

\bibitem{goldenberg2013accurate}
N.~Goldenberg and A.~Wool, ``Accurate modeling of modbus/tcp for intrusion
  detection in scada systems,'' \emph{International Journal of Critical
  Infrastructure Protection}, vol.~6, no.~2, pp. 63--75, 2013.

\bibitem{chakrabarti2010study}
S.~Chakrabarti, M.~Chakraborty, and I.~Mukhopadhyay, ``Study of snort-based
  ids,'' in \emph{Proceedings of the International Conference and Workshop on
  Emerging Trends in Technology}.\hskip 1em plus 0.5em minus 0.4em\relax ACM,
  2010, pp. 43--47.

\bibitem{lin2013adapting}
H.~Lin, A.~Slagell, C.~Di~Martino, Z.~Kalbarczyk, and R.~K. Iyer, ``Adapting
  bro into scada: building a specification-based intrusion detection system for
  the dnp3 protocol,'' in \emph{Proceedings of the Eighth Annual Cyber Security
  and Information Intelligence Research Workshop}.\hskip 1em plus 0.5em minus
  0.4em\relax ACM, 2013, p.~5.

\bibitem{mehra2012brief}
P.~Mehra, ``A brief study and comparison of snort and bro open source network
  intrusion detection systems,'' \emph{International Journal of Advanced
  Research in Computer and Communication Engineering}, vol.~1, no.~6, pp.
  383--386, 2012.

\bibitem{salem2016anomaly}
M.~Salem, M.~Crowley, and S.~Fischmeister, ``Anomaly detection using
  inter-arrival curves for real-time systems,'' in \emph{Real-Time Systems
  (ECRTS), 2016 28th Euromicro Conference on}.\hskip 1em plus 0.5em minus
  0.4em\relax IEEE, 2016, pp. 97--106.

\bibitem{gryllias2012support}
K.~C. Gryllias and I.~A. Antoniadis, ``A support vector machine approach based
  on physical model training for rolling element bearing fault detection in
  industrial environments,'' \emph{Engineering Applications of Artificial
  Intelligence}, vol.~25, no.~2, pp. 326--344, 2012.

\bibitem{durgesh2010data}
K.~S. Durgesh and B.~Lekha, ``Data classification using support vector
  machine,'' \emph{Journal of Theoretical and Applied Information Technology},
  vol.~12, no.~1, pp. 1--7, 2010.

\bibitem{chang2011libsvm}
C.-C. Chang and C.-J. Lin, ``Libsvm: a library for support vector machines,''
  \emph{ACM Transactions on Intelligent Systems and Technology (TIST)}, vol.~2,
  no.~3, p.~27, 2011.

\bibitem{suguna2010improved}
N.~Suguna and K.~Thanushkodi, ``An improved k-nearest neighbor classification
  using genetic algorithm,'' \emph{International Journal of Computer Science
  Issues}, vol.~7, no.~2, pp. 18--21, 2010.

\bibitem{olvera2010review}
J.~A. Olvera-L{\'o}pez, J.~A. Carrasco-Ochoa, J.~F. Mart{\'\i}nez-Trinidad, and
  J.~Kittler, ``A review of instance selection methods,'' \emph{Artificial
  Intelligence Review}, vol.~34, no.~2, pp. 133--143, 2010.

\bibitem{sinwar2014study}
D.~Sinwar and R.~Kaushik, ``Study of euclidean and manhattan distance metrics
  using simple k-means clustering,'' \emph{INTERNATIONAL JOURNAL FOR RESEARCH
  IN APPLIED SCIENC E AND ENGINEERING TECHNOLOGY (IJRASET)}, vol.~2, no.~5, pp.
  270--274, 2014.

\bibitem{hardle2012smoothing}
W.~H{\"a}rdle, \emph{Smoothing techniques: with implementation in S}.\hskip 1em
  plus 0.5em minus 0.4em\relax Springer Science \& Business Media, 2012.

\bibitem{dasgupta2014optimal}
S.~Dasgupta and S.~Kpotufe, ``Optimal rates for k-nn density and mode
  estimation,'' in \emph{Advances in Neural Information Processing Systems},
  2014, pp. 2555--2563.

\bibitem{chauhan2013implementation}
H.~Chauhan and A.~Chauhan, ``Implementation of decision tree algorithm c4. 5,''
  \emph{International Journal of Scientific and Research Publications}, vol.~3,
  no.~10, 2013.

\bibitem{farid2010attacks}
D.~M. Farid, N.~Harbi, E.~Bahri, M.~Z. Rahman, and C.~M. Rahman, ``Attacks
  classification in adaptive intrusion detection using decision tree,''
  \emph{World Academy of Science, Engineering and Technology}, vol.~63, pp.
  86--90, 2010.

\bibitem{borghoff2010hill}
J.~Borghoff, L.~R. Knudsen, and K.~Matusiewicz, ``Hill climbing algorithms and
  trivium,'' in \emph{International Workshop on Selected Areas in
  Cryptography}.\hskip 1em plus 0.5em minus 0.4em\relax Springer, 2010, pp.
  57--73.

\bibitem{moy2010arrival}
M.~Moy and K.~Altisen, ``Arrival curves for real-time calculus: the causality
  problem and its solutions,'' in \emph{International Conference on Tools and
  Algorithms for the Construction and Analysis of Systems}.\hskip 1em plus
  0.5em minus 0.4em\relax Springer, 2010, pp. 358--372.

\bibitem{macfarland2016mann}
T.~W. MacFarland and J.~M. Yates, ``Mann--whitney u test,'' in
  \emph{Introduction to Nonparametric Statistics for the Biological Sciences
  Using R}.\hskip 1em plus 0.5em minus 0.4em\relax Springer, 2016, pp.
  103--132.

\bibitem{shrivastava2011effective}
S.~K. Shrivastava and P.~Jain, ``Effective anomaly based intrusion detection
  using rough set theory and support vector machine,'' \emph{International
  Journal of Computer Applications}, vol.~18, no.~3, pp. 35--41, 2011.

\bibitem{dongarra2013toward}
J.~Dongarra and M.~A. Heroux, ``Toward a new metric for ranking high
  performance computing systems,'' \emph{Sandia Report, SAND2013-4744}, vol.
  312, 2013.

\end{thebibliography}
\bibliographystyle{IEEEtran}



\vfill







\end{document}